%% file: _main.tex
\input{_constants}
\arxiv

\pdfoutput=1
\documentclass[10pt,twocolumn,letterpaper]{article}
\input{cvpr_header}
\begin{document}
\title{UniMuMo: Unified Text, Music and Motion Generation
}
\author{\authorBlock}


\twocolumn[{
    \renewcommand\twocolumn[1][]{#1}
    \maketitle
    \centering
    \vspace{-4em}
    \begin{minipage}{1.00\textwidth}
        \centering
        \includegraphics[trim=000mm 000mm 000mm 000mm, clip=False, width=\linewidth]{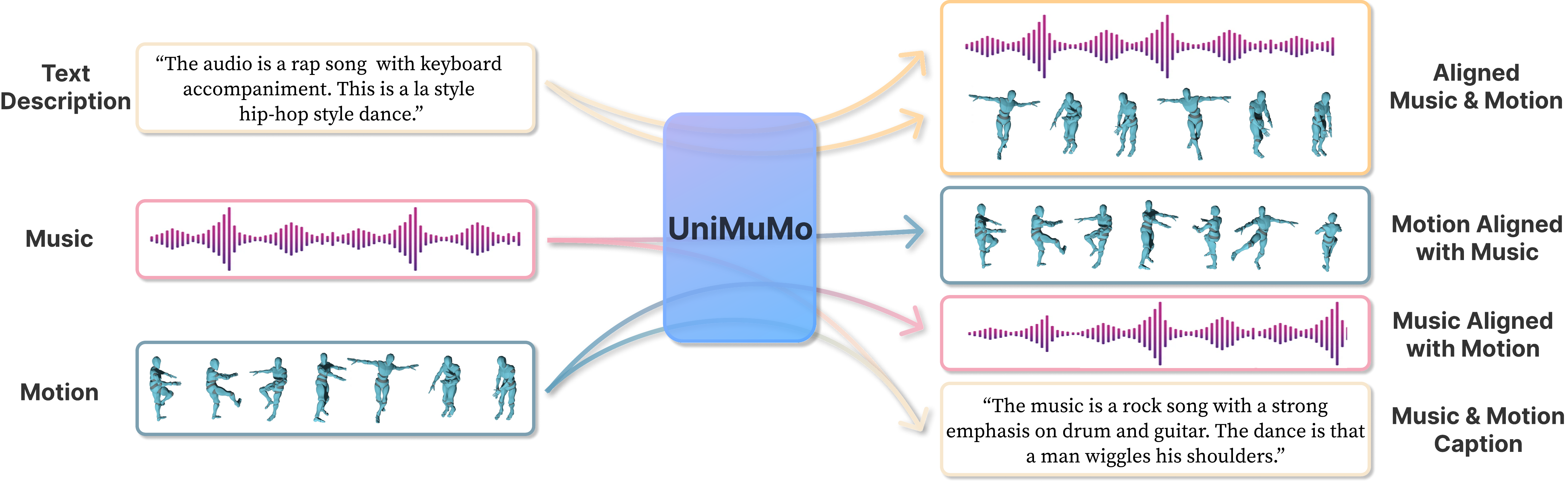}
    \end{minipage}
    \vspace{-3mm}
    \captionsetup{type=figure}
    \captionof{figure}{\small UniMuMo is able to perform generation tasks on any combination of music, motion, and text. The tasks shown in the figure include text-to-aligned-music-motion, music-to-motion, motion-to-music, music-captioning, and motion-captioning.}
    \label{fig:teaser}
    \vspace{3mm}
}]


\input{00_abstract}

\input{01_intro}
\input{02_related}

\input{03_method}

\input{04_experiment}

\input{10_conclusion}

{\small
\bibliographystyle{ieeenat_fullname}
\bibliography{11_references}
}

\ifarxiv \clearpage \appendix \input{12_appendix} \fi

\end{document}

%% file: _constants.tex
\def\paperID{0996}
\def\confName{CVPR}
\def\confYear{2024}

\def\authorBlock{
    Han Yang$^1$ \qquad
    Kun Su$^2$ \qquad
    Yutong Zhang$^3$ \qquad
    Jiaben Chen$^4$ \\
    Kaizhi Qian$^5$ \qquad
    Gaowen Liu$^6$ \qquad
    Chuang Gan$^{4,5}$ \\
    $^1$The Chinese University of Hong Kong \qquad
    $^2$University of Washington \\
    $^3$The University of British Columbia\qquad
    $^4$UMass Amherst \qquad
    $^5$MIT-IBM Watson AI Lab \\
    $^6$Cisco Research \\
    {\tt\small \{email, addresses\}@inst.edu}
}

\newif\ifreview 
\newif\ifarxiv \newcommand{\arxiv}{\arxivtrue}
\newif\ifcamera 
\newif\ifrebuttal 

%% file: cvpr_header.tex
\ifreview \usepackage[review]{cvpr} \fi
\ifarxiv \usepackage[pagenumbers]{cvpr} \fi
\ifrebuttal \usepackage[rebuttal]{cvpr} \fi
\ifcamera \usepackage{cvpr} \fi

\input{_macros}  

\usepackage{xr-hyper}

\makeatletter
\newcommand*{\addFileDependency}[1]{
  \typeout{(#1)}
  \@addtofilelist{#1}
  \IfFileExists{#1}{}{\typeout{No file #1.}}
}

\makeatother

\definecolor{cvprblue}{rgb}{0.21,0.49,0.74}
\usepackage[pagebackref,breaklinks,colorlinks,citecolor=cvprblue]{hyperref}
\usepackage[capitalize]{cleveref}
\crefname{section}{Sec.}{Secs.}
\crefname{table}{Table}{Tables}
\crefname{figure}{Fig.}{Figs.}

%% file: _macros.tex
\usepackage{graphicx}	
\usepackage{amsmath}	
\usepackage{amssymb}	
\usepackage{booktabs}
\usepackage{times}
\usepackage{float}
\usepackage{microtype}
\usepackage{epsfig}
\usepackage[table,xcdraw,dvipsnames]{xcolor}
\usepackage{caption}
\usepackage{float}
\usepackage{placeins}
\usepackage{color, colortbl}
\usepackage{stfloats}
\usepackage{enumitem}
\usepackage{tabularx}
\usepackage{xstring}
\usepackage{multirow}
\usepackage{xspace}
\usepackage{url}
\usepackage{subcaption}
\usepackage{xcolor}
\usepackage[hang,flushmargin]{footmisc}
\usepackage{bbm}
\ifcamera \usepackage[accsupp]{axessibility} \fi

% ...

\ifarxiv  \fi

\newcommand{\R}[1]{{%
    \textbf{%
        \ifstrequal{#1}{1}{\textcolor{red}{R#1}}{%
        \ifstrequal{#1}{2}{\textcolor{blue}{R#1}}{%
        \ifstrequal{#1}{3}{\textcolor{magenta}{R#1}}{%
        \ifstrequal{#1}{4}{\textcolor{teal}{R#1}}{%
                           \textcolor{cyan}{R#1}%
        }}}}%
    }%
}}

%% file: 00_abstract.tex
\begin{abstract}
We introduce UniMuMo, a unified multimodal model capable of taking arbitrary text, music, and motion data as input conditions to generate outputs across all three modalities. To address the lack of time-synchronized data, we align unpaired music and motion data based on rhythmic patterns to leverage existing large-scale music-only and motion-only datasets. By converting music, motion, and text into token-based representation, our model bridges these modalities through a unified encoder-decoder transformer architecture. To support multiple generation tasks within a single framework, we introduce several architectural improvements. We propose encoding motion with a music codebook, mapping motion into the same feature space as music. We introduce a music-motion parallel generation scheme that unifies all music and motion generation tasks into a single transformer decoder architecture with a single training task of music-motion joint generation. Moreover, the model is designed by fine-tuning existing pre-trained single-modality models, significantly reducing computational demands. Extensive experiments demonstrate that UniMuMo achieves competitive results on all unidirectional generation benchmarks across music, motion, and text modalities. Quantitative results are available in \href{https://hanyangclarence.github.io/unimumo_demo/}{https://hanyangclarence.github.io/unimumo\_demo/}.
\end{abstract}

%% file: 01_intro.tex
\section{Introduction}
\label{sec:intro}

Music and body movements are synchronized and inseparable. The beat and metrical structures in rhythm encourage the spontaneous coordination of body motion with music~\cite{large2000synchronizing}, activating the motor-related areas of human brains~\cite{keller2009musical}. Dance particularly exemplifies this connection through choreography that aligns with the music's rhythm, melody and emotion. Meanwhile, even though most people are not professional musicians or dancers, they often interpret music and dance using simple, natural language. This descriptive text serves as a vital bridge  between understandable ideas and abstract concepts in music and motion.

The synergy between music, motion, and text provides a natural motivation to create a model capable of understanding and creating contents across all these modalities.  Moreover, building a framework that can flexibly generate music, motion, and text in arbitrary combinations is crucial for real-world applications, even though existing models already achieve impressive results in unidirectional generation tasks such as text-to-music~\cite{copet2023simple}, music-to-motion~\cite{tseng2023edge}, motion-to-music~\cite{zhu2022quantized} and motion-to-text~\cite{jiang2023motiongpt}. In the real world, there is a demand for diverse generative abilities, and more complex generation tasks may be necessary, such as creating dance sequences based on both music and textual descriptions. Training individual models for each unique combination, although potentially yielding better output quality, would significantly increase training costs, deployment efforts and storage requirements. Thus, a unified model that supports all combinations of conditioning and generation tasks, rather than a collection of separate models or training adapters to incorporate individual models, offers a more cost-effective solution. To this end, we introduce a novel task of dynamically generating music, motion, and text in a multitude of combinations unifiedly. As demonstrated in Fig.~\ref{fig:teaser}, this task is designed to handle diverse generative scenarios, ranging from text-to-music, text-to-motion, to more complex combinations like text-to-music-plus-motion or music-plus-text-to-motion.

However, the task could be challenging, especially in two aspects:  i) the lack of comprehensive datasets that include all three modalities - music, motion, and text - limits the development of a general and unified model. While there are individual datasets for music-only~\cite{santana2020music4all}, motion-only~\cite{AMASS:ICCV:2019}, music to motion~\cite{li2021learn} and text to motion~\cite{Guo_2022_CVPR}, a holistic and large-scale dataset that encompasses all three modalities still remains absent; 
ii) designing a unified architecture that supports both the conditioning  and generation of all three modalities is challenging, mainly due to the significant differences between the neural representations for the three modalities and the multiplicity of desired generation tasks.  

To address the first challenge of lacking paired data, we propose to align unpaired music and motion sequences based on their rhythmic patterns. Specifically, we extract both music beats and motion visual beats, then employ dynamic time warping to find the alignment and warp the motion sequence to adjust the motion visual beats to match the music beats. We found that such augmentation is accurate and efficient. With the augmented synchronized music-motion data, we can utilize existing music and motion datasets to train our unified generative model. 
Additionally, we construct text descriptions from music and motion metadata using a mixture of template filling, large language model generation and music-based language model generation, striking a balance between diversity, language fluency and description accuracy.

To overcome the second challenge, we propose a novel framework, UniMuMo, to unify the generation of different modalities. Our pipeline consists of three main stages: a music-motion joint tokenizer that encodes music and motion sequences into discrete representations within the same space, a music-motion transformer-decoder model trained on the task of music-motion joint generation, and a music-motion captioner that generates text descriptions from music and motion features. 
In the first stage, we bridge the modality gap between music and motion by mapping motion into the music feature space. Specifically, instead of using separate Vector-Quantized Variational Autoencoders (VQ-VAE) to quantize music and motion sequences, we encode motion with the codebook of a pre-trained music VQ-VAE, namely Encodec~\cite{defossez2022high}. This design facilitates the unification of music and motion within the same generative framework in the subsequent stage.
In the second stage, we train a unified music and motion generative model with a novel task of music-motion joint generation from text conditions. To enable the mutual conditioning of music and motion, and unlock the music-to-motion and motion-to-music generation capabilities, we introduce a novel music-motion parallel generation scheme, where we perform two mutually conditioned streams of autoregressive generation of aligned music and motion simultaneously. With the reuse of Encodec and joint encoding of motion in the previous stage, the current stage can be effectively achieved by fine-tuning the pre-trained text-to-music model associated with Encodec, namely MusicGen~\cite{copet2023simple}, equipping it with additional motion conditioning and generation capabilities while maintaining its music generation capabilities.
In the third stage, we fine-tune a T5 decoder for music and motion captioning tasks, using the features extracted by the music-motion decoder trained in stage 2. To transform the decoder into an effective feature extractor, we replace its causal self-attention layers with trainable full self-attention layers, and fine-tune them together with the T5 decoder on music and motion captioning tasks.
Extensive experiments demonstrate that UniMuMo achieves competitive performance across all unidirectional generation tasks in music, motion, and text when compared with existing state-of-the-art models, demonstrating the effectiveness and versatility of our approach.

Our work offers significant advancements in multimodal generative research, summarized as follows:
\begin{itemize}
    \item To the best of our knowledge, this is the first unified framework capable of arbitrarily generating content across music, motion, and text.
    \item To address the shortage of paired multimodal data, we augment and enrich existing large-scale datasets with music-motion data alignment and text augmentations.
    \item We propose a novel joint codebook for encoding music and motion sequences, along with a music-motion parallel generation scheme, facilitating multiple generation tasks within a single architecture.
    \item  Our framework achieves results comparable to SOTAs across all generation tasks in music, motion, and text.
\end{itemize}

%% file: 02_related.tex
\section{Related Work}
\label{sec:related}

\noindent\textbf{Text to Music}.
Text-conditioned music generation has been widely studied in recent years. There are two main branches: diffusion-based and transformer-based. For diffusion-based models, Riffusion~\cite{Forsgren_Martiros_2022} uses a latent text-to-image diffusion model to generate spectrograms, which are then converted into audio clips; Mousai~\cite{schneider2023mo} proposes training a diffusion model in the latent space of a diffusion autoencoder; Noise2Music~\cite{huang2023noise2music} introduces a cascade of diffusion models that first generates the audio in a coarse form and then progressively refine it. AudioLDM~\cite{liu2023audioldm2} proposes to train a latent diffusion model using CLAP~\cite{wu2023large} embeddings, a language-audio joint representation, for text conditioning. 
For transformer-based models, MusicLM~\cite{Agostinelli2023MusicLMGM} proposes to encode music into high-level "semantic tokens" and low-level "acoustic tokens", and use a cascade of transformer decoders to generate the two levels stage by stage. MusicGen~\cite{copet2023simple} leverages a single-stage transformer decoder to model the hierarchical music tokens directly.

\noindent\textbf{Music to Text}.
Several models have been proposed for audio captioning. 
WAC~\cite{kadlvcik2023whisper} proposes to transfer a pre-trained speech-to-text Whisper model to the music captioning task. LTU~\cite{gong2023listen} takes the concatenated music embeddings and text embeddings as input to a large language model and directly trains caption generation using language modeling objectives. LP-MusicCaps~\cite{doh2023lp} uses a transformer encoder-decoder structure, where the music spectrogram is first encoded by the encoder and then cross-attended by the decoder for text generation. MU-LLaMA~\cite{liu2023music} leverages a frozen LLaMA~\cite{touvron2302llama} and fine-tunes a Music Understanding Adapter to fuse music features into the LLaMA model.

\noindent\textbf{Music to Motion}. 
Most of the works on music-conditioned dance generation are based on transformers.
Several approaches~\cite{li2021ai, fan2022bi, pu2022music} adopt similar structures that first use a music transformer encoder and a motion transformer encoder to encode music and initial motion into representations separately, and then employ a transformer decoder for cross-modal fusion and motion generation.
Bailando~\cite{siyao2022bailando} proposes to train a transformer on motion features encoded by a choreographic memory module, which is the codebook of a motion VQ-VAE. 
Besides autoregressive transformers, EDGE~\cite{tseng2023edge} adopts a transformer-based diffusion model capable of both dance generation and editing.

\noindent\textbf{Motion to Music}.
Most of the relevant works focus on generating corresponding music from video input. Foley Music~\cite{gan2020foley} focuses on generating music for videos of people playing instruments, and uses Musical Instrument Digital Interface
(MIDI) to bridge the gap between body key points and the final music. Similarly, RhythmicNet~\cite{su2021does} extends the scenarios to arbitrary motion videos by first estimating visual rhythm and conditionally generating drum and piano music. 
Dance2Music~\cite{aggarwal2021dance2music} encodes a dance similarity matrix with CNN and predicts the next note with an LSTM autoregressively. CDCD~\cite{zhu2022discrete} proposes a single-stage method that uses a discrete latent diffusion model to generate music spectrograms conditioned on video features. D2M-GAN~\cite{zhu2022quantized} proposes a GAN-based model to generate the music tokens based on video and pose features.

\noindent\textbf{Text to Motion}. 
Text-to-motion approaches can be mainly categorized into transformer-based and diffusion-based. 
Transformer-based models mainly work with motion tokens generated by a motion VQ-VAE. 
TM2T~\cite{guo2022tm2t} regards text-to-motion and motion-to-text as machine translation tasks and trains a transformer encoder-decoder to perform bidirectional translation uniformly. T2M-GPT~\cite{zhang2023t2m} adopts a GPT-based model to directly generate motion tokens from text conditions.  MotionGPT~\cite{jiang2023motiongpt} proposes incorporating motion tokens into natural language tokens and performing language modeling on both text and motion.
For the diffusion-based model, MotionDiffuse~\cite{zhang2022motiondiffuse} and MDM~\cite{tevet2022human} directly adopt diffusion model while MLD~\cite{chen2023executing} adopt latent diffusion on motion generation.

\noindent\textbf{Motion to Text}.
Motion-to-text generation is usually regarded as a sequence-to-sequence translation task. Various methods adopts different sequence-to-sequence models, such as RNN~\cite{plappert2018learning}, recurrent autoencoder~\cite{yamada2018paired}, sequence generative adversarial nets~\cite{goutsu2021linguistic} and transformer~\cite{guo2022tm2t, jiang2023motiongpt}.

%% file: 03_method.tex
\section{Text-Music-Motion Aligned Data Generation}
\label{sec:data_generation}

To model arbitrary generation across music, motion, and text, we propose to expand existing music and motion datasets by aligning motion with music and synthesizing textual descriptions. The data generation pipeline includes four major steps: 1) music beat detection, 2) visual beat detection, 3) music-motion alignment, and 4) text description synthesis.

\noindent\textbf{Music Beat Detection}. 
We estimate music beats from a music waveform $Y \in \mathbb{R}^{T_w}$, where $T_w$ represents the number of samples, using a Bidirectional-LSTM-based model from ~\cite{chiu2021drum}. This model performs beat tracking on extracted drum features and non-drum features separately, then aggregates the results with a learnable fuser. We manually evaluate the accuracy of this beat tracking model and find that it performs well in most test cases, outperforming the beat tracking methods in the Librosa API~\cite{mcfee2015librosa}. The resulting music beats are represented as a binary sequence $B_m \in \mathbb{R}^{T_w}$, where each frame  is marked as `beat' or `non-beat.'

\noindent\textbf{Visual Beats Detection}. 
Given a 3D motion sequence $M \in \mathbb{R}^{T_m \times J \times 3}$ where $T_m$ represents the number of frames, $J$  the number of joints, and the last dimension indicates $x,y,z$ coordinates, we obtain visual beats in three steps. In the first stage, we calculate the motion directogram~\cite{davis2018visual}, a 2D matrix that factors motion into different motion angles, similar to how an audio spectrogram factors sound amplitude into different frequencies. Specifically, we first compute the first-order difference of the motion sequence $\Delta M_t = M_t - M_{t-1}$. Based on its motion angle, we assign the motion magnitude of every joint into one of the bins in $2\pi /N_{\text{bins}}$. The motion directogram $M_d(t, \theta)$ is obtained by summing the motion magnitudes of each bin:
$M_d(t, \theta) = \sum_j \Delta M_t(j)   \mathbf{1}_{\theta}(\angle{M_t(j)})$, where $\mathbf{1}_{\theta}(\phi) = 1 \text{ if } |\theta - \phi| \leq 2\pi/N_{\text{bins}} \text{ else } 0$.
In the second stage, we convert the motion directogram to the kinematic offset $M_k$, which represents the motion changes, similar to the onset envelope in an audio spectrogram. We first obtain motion flux $M_f$, which represents the deceleration in various directions, by computing the negative first-order difference of the directogram $\Delta M_d$. We then average each frame of $M_f$ and filter the top $1\%$ peaks to obtain kinematic offset $M_k$.
In the last stage, we use dynamic programming to compute the visual beats by designing an objective function that selects strong visual changes from kinematic offsets and encourages equal-spacing beats. More details can be found in Appendix~\ref{appendix:visual beat detections}. The final visual beats are also represented as a binary sequence $B_v \in \mathbb{R}^{T_m}$, where each frame is marked as `beat' or `non-beat'.

\noindent\textbf{Music-Motion Alignment.}
We apply dynamic time warping to determine the optimal matching between music beats $B_m$ and visual beats $B_v$, finding the alignment even though the duration of these two binary sequences could be different. Finally, we warp motion sequences by interpolating according to the warping curve to obtain aligned music-motion pairs. The reason for warping motion to match music, rather than the reverse, is that music beats tend to be steady, so warping music could result in perceptually unacceptable changes. More details can be found in  Appendix~\ref{appendix:music motion alignment}.

\noindent \textbf{Text Description Synthesis.}
To compensate for the absence of text descriptions in our used datasets, we employ two methods for captions synthesis: (1) using Music Understanding Language Model to generate caption directly from audio; and (2) using Large Language Model to synthesize captions from metadata (genre, tempo, \etc), striking a balance between musical accuracy and diversity. Examples and more details are shown in Appendix~\ref{appendix:text constructions}.

\section{UniMuMo Framework}
\label{sec:method}
\begin{figure*}[htpb]
    \centering
    \includegraphics[width=\textwidth]{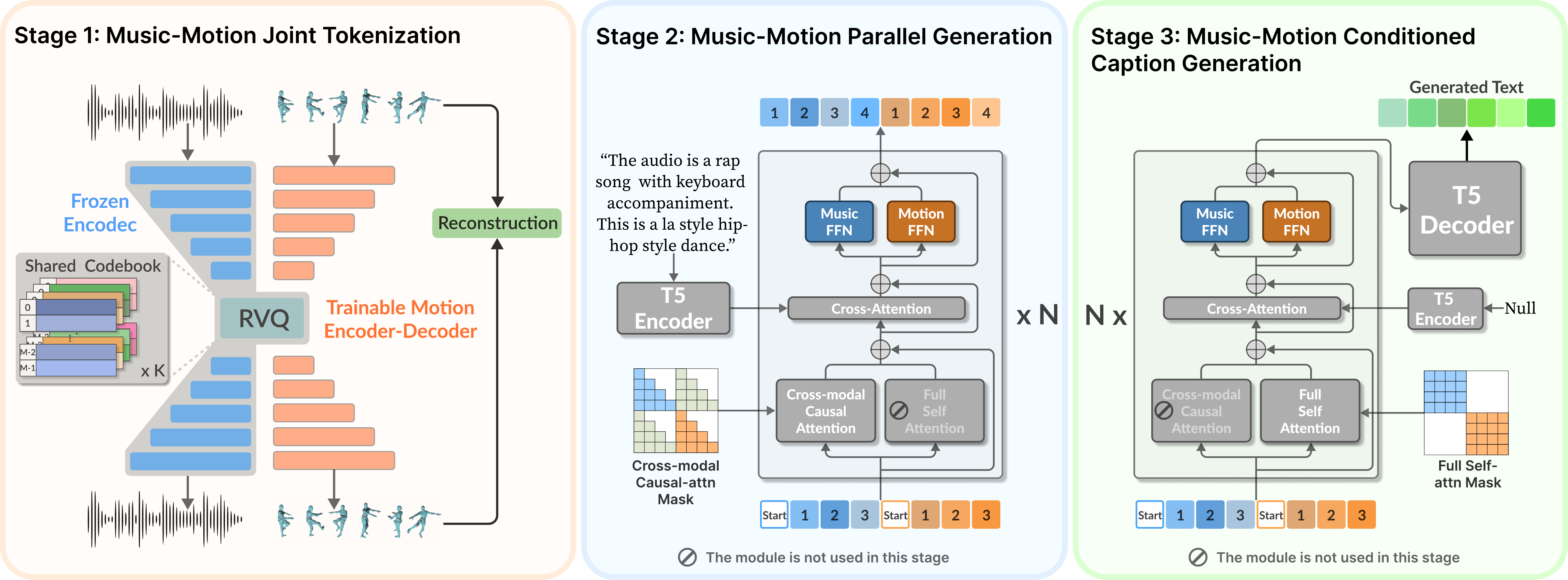}
    \caption{\small Overview: The training of UniMuMo consists of three stages: In stage 1, we train a motion RVQ-VAE using the frozen codebook from a pre-trained music RVQ-VAE to encode motion into the same space as music. In stage 2, we fine-tune a pre-trained music transformer decoder model on the text-to-music-motion task using the music-motion parallel generation scheme. In stage 3, we fine-tune a T5 decoder for music-motion captioning using the previous music-motion decoder as a feature extractor.
    }
    \label{fig:overview}
\vspace{-4mm}
\end{figure*}
UniMuMo consists of three training stages to enable arbitrary generation between music, motion, and text.
In stage 1, we encode aligned music and motion data into discrete tokens. To efficiently bridge the gap between the two modalities, we propose to use a frozen pre-trained audio tokenizer Encodec~\cite{defossez2022high} and train a motion tokenizer that reuses the same residual codebooks of the audio tokenizer.
In stage 2, we fine-tune a state-of-the-art text-to-music transformer decoder~\cite{copet2023simple} by conducting the task of generating music and motion tokens simultaneously with music and motion text descriptions. At the inference stage, we can perform parallel generation to unlock applications of music and motion generation. In stage 3, we treat the pre-trained music-motion decoder model in stage two as a feature extractor and fine-tune a T5 decoder on language modeling task for music and motion captioning. An overview of the UniMuMo framework is shown in Figure~\ref{fig:overview}.

\subsection{Stage 1. Music and Motion Joint Tokenization}\label{sec/method/tokenization}
While existing tokenization approaches can faithfully reconstruct the music or motion individually, the correlations between the two modalities become intricate in distinct spaces. Therefore, directly applying them in the unified generation framework poses challenges. Besides, a music tokenizer usually requires more training resources and time to achieve high-quality reconstruction than a motion tokenizer. 
Inspired by these facts, we introduce an efficient and effective way to encode music and motion into a joint latent space. We propose using a pre-trained audio tokenizer, Encodec~\cite{defossez2022high}, and training a new motion encoder-decoder. The motion encoder encodes the motion into the same embedding space as the music and reuses the frozen music Residual Vector Quantizers (RVQ) to discretize the motion into tokens. From these tokens, the motion decoder can decode to reconstruct the motion. Given the higher complexity and richer information in music compared to motion, the learned music codebook is theoretically capable of encoding motion.

Specifically, given a waveform $Y \in \mathbb{R}^{T \cdot f_w}$ with $T$ the audio duration and $f_w$ the sample rate, Encodec first encodes it into a continuous tensor of $X_{\text{music}} \in \mathbb{R}^{d \times T \cdot f_r}$, where $f_r \ll f_w$ is the frame rate of the residual codebook and $d$ is the dimension of codebook entries. $X_{\text{music}}$ is then quantized by the RVQ into music tokens $Q_{\text{music}} \in \{1, \dots, M\}^{K \times T \cdot f_r}$, where $K$ is the number of RVQ and $M$ is the number of codebook entries. For an aligned motion sequence of the same duration $M \in \mathbb{R} ^ {d_m \times T \cdot f_m}$ with frame rate $f_m$ and feature dimension $d_m$, our motion encoder encodes it into $X_{\text{motion}} \in \mathbb{R}^{d \times T \cdot f_r}$, the same shape as $X_{\text{music}}$, which is then tokenized by the same RVQ into motion tokens $Q_{\text{motion}} \in \{1, \dots, M\}^{K \times T \cdot f_r}$. The motion decoder decodes the motion feature after RVQ, resulting in $\hat{M}$. The motion encoder-decoder is trained by minimizing the motion reconstruction loss together with a commitment loss $\mathcal{L}_{\text{commit}}$ from the codebook: 
\vspace{-2mm}
\begin{equation}
    \mathcal{L}_{\text{total}} = \frac{1}{|\mathcal{D}|} \sum_{M \in \mathcal{D}}
    (\| M - \hat{M} \|_2 + \lambda \mathcal{L}_{\text{commit}})
\vspace{-2mm}
\end{equation}
where $\mathcal{D}$ is the motion dataset and $\lambda$ controls the strength of the commitment loss. Empirically, $\lambda$ is set to 0.02.

With this design, the music-motion joint tokenization can effectively learn multimodal correlations by mapping motion features into the same space as music, without the need to train another computationally heavy music autoencoder.
Moreover, it enables direct use the text-to-music model associated with Encodec as an initialization for the following music-motion decoder model, significantly reducing training costs and enhancing the performance.
Experimentally, such feature alignment is crucial to learning the joint generation of music and motion within a single transformer model.

\subsection{Stage 2. Music and Motion Generation from Text}\label{sec/method/training}
In this stage, we modify and fine-tune an existing state-of-the-art text-to-music model with the music and motion tokens extracted from Stage 1, enabling it to handle all tasks related to music and motion generation, such as text-to-music-motion and motion-to-music.
In particular, we employ MusicGen~\cite{copet2023simple}, an open-source, single-stage transformer decoder model that can generate multi-level music tokens with a specific codebook interleaving pattern. Following their practice, we apply the delay pattern for both music and motion tokens, utilize a T5 encoder for encoding text descriptions, and adopt cross-attention to incorporate text conditioning features into the transformer decoder.

To enable the autoregressive generation of music and motion within a unified  framework, we propose training on the task of music-motion joint generation, together with a novel parallel generation scheme, where two streams (\ie, music and motion) of predict-next-token generation are conducted simultaneously, with each stream conditioned on each other. Specifically, given the music tokens $Q_{\text{music}}$ and motion tokens $Q_{\text{motion}}$ with the same shape $K \times S$ where $S = T \cdot f_r$ is the sequence length, we first transform them with delay pattern~\cite{copet2023simple} into $Q_{\text{music}}'$ and $Q_{\text{motion}}'$ respectively, resulting shape $K \times S'$, where $S' = S + K - 1$. We then concatenate them in time dimension into $Q_{\text{input}}$ of the shape $K \times 2S'$ as the input to the transformer decoder. The model's output is transformed back to the normal pattern for loss calculation. 
Training on music-motion joint generation, we adopt the predict-next-token objectives for both music and motion tokens in each forward pass:

\begin{equation}
\footnotesize
\begin{split}
    \mathcal{L} =& - \frac{1}{|\mathcal{D}|} \sum_{Q \in \mathcal{D}}  \left \{ \mu \cdot \sum_{t = 1}^{S}\log \mathbb{P}\left [Q_t^{\text{music}} | Q_{<t}^{\text{music}}, Q_{<t}^{\text{motion}}\right] \right. \\ 
  &\left. + (1 - \mu) \cdot \sum_{t = 1}^{S}\log \mathbb{P}\left[Q_t^{\text{motion}} | Q_{<t}^{\text{music}}, Q_{<t}^{\text{motion}}\right] \right\}
\end{split}
\end{equation}

\noindent where $\mu$ balances between music loss and motion loss, and $\mathbb{P}$ denotes predict-next-token probability of the model. Empirically, $\mu$ is set to 0.85. To enable the parallel autoregressive generation, we apply a cross-modal causal attention mask, as shown in Stage 2 of Figure~\ref{fig:overview}. The causal attention mask is of shape $2S' \times 2S'$, each quarter of which is an $S' \times S'$ lower triangular matrix, allowing music and motion tokens to have both cross-modal and uni-modal causal attention. A further illustration of the strategy can be found in Appendix~\ref{appendix:parallel generation}. 

With the above construction, the model can perform parallel sampling during inference, enabling the prediction of the next token for both music and motion concurrently:

{\footnotesize
\begin{align}
&\hat{Q}_t^{\text{music}} = \underset{i \in M}{\text{argmax}} \mathbb{P}[Q_{t,i} ^{\text{music}} | \hat{Q}_{<t}^{\text{music}}, \hat{Q}_{<t}^{\text{motion}}] \\
&\hat{Q}_t^{\text{motion}} = \underset{i \in M}{\text{argmax}} \mathbb{P}[Q_{t, i}^{\text{motion}} | \hat{Q}_{<t}^{\text{music}}, \hat{Q}_{<t}^{\text{motion}}]
\end{align}
}

\noindent where $M$ is the codebook size. With this sampling strategy, we can conduct the joint generation of music and motion under text conditions.
Additionally, it facilitates zero-shot music-to-motion and motion-to-music generation. For example, given a music sequence $Q_{1:S}^{\text{music}}$, an aligned motion sequence can be autoregressively sampled by

\begin{equation}
\footnotesize
    \hat{Q}_t^{\text{motion}} = \underset{i \in M}{\text{argmax}} \mathbb{P}[Q_{t, i}^{\text{motion}} | Q_{<t}^{\text{music}}, \hat{Q}_{<t}^{\text{motion}}] 
\end{equation}

\noindent An illustration of the sampling process can also be found in Appendix~\ref{appendix:parallel generation}.

Considering the inherent differences between music and motion, we further introduce the following changes to the pre-trained MusicGen to alleviate the mutual interference between the two modalities. First, we add another trainable embedder for motion tokens, with which the model can learn to differentiate the two modalities. Second, to ensure the temporal parallelism, we add positional encodings $\{E_1, E_2, \dots, E_{S'}\}$ to music and motion separately, instead of using a holistic positional encoding of length $2S'$. Third, inspired by the idea of Mixture of Experts (MoE), we introduce an additional feed-forward network (FFN) for motion in each transformer layer. As shown in Fig.~\ref{fig:overview}, in each forward pass, the first half of the feature (\ie, music features) is processed by the music FFN, and the second half (\ie, motion features) by the motion FFN. Fourth, we add a new motion classification head at the end of the network to distinguish motion code prediction from music code prediction. Note that for the new modules introduced above, we initialize the motion embedder and FFNs with the corresponding components from the pre-trained MusicGen. With a joint motion VQ-VAE trained in Stage 1, such initialization ensures that music features are not confused by uninitialized motion features at the beginning of training, allowing the music generation capability to be better preserved.

Following MusicGen, text conditioning is added with cross-attention. In the framework of music-motion joint generation, we add the text condition of two modalities independently. We first encode music descriptions and motion descriptions separately into features and apply classifier-free guidance dropout independently. Then, during cross-attention on text conditions, we specialize the attention mask to allow music features to attend only to music conditions and motion features to attend only to motion conditions. 

By fine-tuning the model on the music-motion dataset with the above settings, we find that the model learns to generate motion in parallel with music quickly while still keeping its music-generation ability. 
With a single training task of music-motion joint generation, various applications could be achieved in a zero-shot fashion, including text-to-music, text-to-motion, music-to-motion, motion-to-music, motion-text-to-music, \etc.

\subsection{Stage 3. Music and Motion Captioning}\label{sec/method/captioning}
The final stage is for caption generation, where we treat the fine-tuned music-motion decoder in the previous stage as a feature encoder for music and motion, and fine-tune another T5 decoder to generate captions for music and motion.

However, using the music-motion decoder directly as a feature extractor brings challenges. Firstly, the self-attention in the decoder is done causally, which is inadequate for capturing rich music and motion features. Secondly, since the input of the model is the concatenation of music and motion,  we are limited to input music-motion pairs for captioning, which is inflexible.

To address these issues, we introduce a trainable full self-attention module, initialized with the trained cross-modal causal attention module, as shown in Fig.~\ref{fig:overview}, Stage 3. Inspired by BLIP~\cite{li2022blip}, which claims that the major difference between transformer encoders and decoders lies in the self-attention layers, with embedding layers and FFNs functioning similarly, we therefore fine-tune only the newly introduced full self-attention modules together with the T5 decoder on caption generation task, keeping the rest of the music-motion decoder unchanged. Considering that captions of music and motion are independent, we remove the cross-attention areas on the attention mask.

In practice, we first randomly mask the entire music or motion tokens as empty, and concatenate them together as input   $Q_{\text{input}}$. This allows us to conduct music or motion captioning independently.
Next, we forward it through the music-motion decoder with a null condition, where full self-attention is applied. We then take the output of the last hidden layer of the model as the feature, which is cross-attended by the T5 text decoder. We fine-tune the model with the language modeling task, and the generation target is either music caption or motion caption, depending on the input masking.

%% file: 04_experiment.tex
\section{Experiment}
\label{sec:experiment}

\subsection{Effectiveness of Music-Motion Alignment}
To quantitatively evaluate the beat alignment results, we calculate the mean L1 distance between each music beat and its nearest visual beat before and after the alignment. Specifically, we randomly sample 300 music clips, each 10 seconds long, pair each with a random motion sequence, and then calculate the score before and after aligning them using the algorithms we introduced. The mean L1 distance decreases from \textbf{6.34} to \textbf{1.78}, demonstrating the overall effectiveness of our alignment algorithms. We further conduct a user study, the results of which are in Appendix~\ref{appendix:user_study}.

\subsection{Evaluations}
We conduct extensive evaluations of our model across various tasks and metrics. More implementation details about hyperparameter choices, dataset, metrics and training/evaluation setups are in Appendix~\ref{appendix:evaluation details}.

\noindent 
\begin{minipage}{\linewidth} 
  \small
  \centering 
  \resizebox{\linewidth}{!}{
  \scalebox{0.9}{
  \begin{tabular}{l|ccc}
    \hline \hline
    Models  & \multicolumn{1}{c}{FAD$_{\text{VGG}}\downarrow$} & \multicolumn{1}{c}{KL$\downarrow$} & \multicolumn{1}{c}{CLAP$\uparrow$} \\ \hline \hline
    Riffusion~\cite{Forsgren_Martiros_2022} & 14.8 & 2.06 & 0.19 \\
    Mubert~\cite{mubert2022}& 9.6 & 1.58 & - \\
    Mousai~\cite{schneider2023mo} & 7.5 & 1.59   & 0.23  \\
    MusicLM  (860M)~\cite{Agostinelli2023MusicLMGM}  & \underline{4.0}    & \underline{1.31}  & -   \\
    MusicGen (300M)~\cite{copet2023simple} & 4.9  & 1.42  & \underline{0.27}   \\
    AudioLDM 2-Full (346M)~\cite{liu2023audioldm2} & \textbf{3.13}  & \textbf{1.17}& \textbf{0.38}\\ \hline
    Ours (300M)   & 5.93    & 1.99    & \underline{0.27}    \\ \hline
    MusicGen (fine-tuned on our data) & 5.81 & 1.97 & 0.28 \\ 
    Ours (trained on data with vocals) & 4.11 & 1.95 & 0.29 \\  \hline  \hline
    \end{tabular}
  }
  }
  \vspace{-3mm}
  \captionof{table}{\small Comparison of text-to-music generation on MusicCaps. \textbf{Bold} and \underline{underlined} results are the best and second-best results.} 
  \label{table:text-to-music}
\end{minipage}%
\vspace{2mm}
\begin{minipage}{\linewidth} 
  \small
  \centering 
  \scalebox{0.8}{
  \begin{tabular}{l|cc}
    \hline \hline
    Models      & Beats Coverage$\uparrow$ & Beats Hit$\uparrow$  \\ \hline \hline
    Dance2Music~\cite{aggarwal2021dance2music} & 83.5           & 82.4          \\
    Foley Music~\cite{gan2020foley} & 74.1           & 69.4          \\
    CMT~\cite{di2021video}         & 85.5           & 83.5           \\
    D2M-GAN~\cite{zhu2022quantized}     & 88.2           & 84.7          \\
    CDCD~\cite{zhu2022discrete}     & \textbf{93.9}          & \textbf{90.7}          \\\hline
    Ours        & \underline{93.0}           & \underline{88.4}         \\ \hline \hline
  \end{tabular}
  }
  \vspace{-3mm}
  \captionof{table}{\small Comparison of motion-conditioned music generation on AIST++.} 
  \label{table:motion-to-music}
\end{minipage}


\noindent 
\begin{minipage}{\linewidth} 
  \small
  \centering 
  \scalebox{0.8}{
  \begin{tabular}{l|ccc}
    \hline \hline
    Models     & Dist$_{\text{k}}\rightarrow$  & Dist$_{\text{g}}\rightarrow$   & Beat Align.$\uparrow$   \\ \hline \hline
    Real         & 10.61   & 7.48    & 0.24          \\ \hline
    Bailando~\cite{siyao2022bailando}  & 7.92    & \underline{7.72}    & 0.23        \\
    FACT~\cite{li2021ai}    & 10.85   & 6.14    & 0.22        \\
    EDGE~\cite{tseng2023edge}    & \textbf{10.58}   & \textbf{7.62}    & \textbf{0.27}        \\ \hline
    Ours (music conditioned)      & \underline{10.68}   & 10.35    & \underline{0.24}        \\
    Ours (text conditioned)   & 9.14   & 9.37    & 0.25        \\\hline \hline
  \end{tabular}
  }
  \vspace{-3mm}
  \captionof{table}{\small Comparison of music-conditioned and text-conditioned dance generation.} 
  \label{table:motion_generation}
\end{minipage}%
\hfill 
\begin{minipage}{\linewidth} 
  \small
  \centering 
  \scalebox{0.8}{
  \begin{tabular}{l|cccc}
    \hline \hline
    Models       & Bleu$\uparrow$   & Meteor$\uparrow$   & Rouge$\uparrow$   & BertScore$\uparrow$ \\ \hline \hline
    LTU~\cite{gong2023listen}          & 0.238 & 0.250 & 0.332 & 0.876  \\
    LP-MusicCaps~\cite{doh2023lp} & 0.165 & 0.202 & 0.281 & 0.879  \\
    MU-LLaMA~\cite{liu2023music}     & \underline{0.238} & \textbf{0.354} & \textbf{0.475} &  \textbf{0.913}  \\ \hline
    Ours         & \textbf{0.261}     & \underline{0.291}     & \underline{0.369}     & \underline{0.892}      \\ \hline \hline
  \end{tabular}
  }
  \vspace{-3mm}
  \captionof{table}{\small Comparison of music captioning on MusicQA dataset.} 
  \label{table:music-to-text}
\end{minipage}
\vspace{1mm}

\begin{table*}[htbp]
\small
\centering
\setlength{\tabcolsep}{1.5mm}{
\scalebox{0.85}{
\begin{tabular}{l|cccccccc}
\hline \hline
\multirow{2}{*}{Methods} & \multicolumn{2}{c}{R-Precision$\uparrow$} & \multirow{2}{*}{MMDist$\downarrow$} & \multicolumn{2}{c}{Bleu$\uparrow$} & \multirow{2}{*}{ROUGE-L$\uparrow$} & \multirow{2}{*}{Cider$\uparrow$} & \multirow{2}{*}{BertScore$\uparrow$} \\ \cline{2-3} \cline{5-6}
                         & Top1           & Top3          &                         & @1          & @4         &                          &                        &                            \\ \hline \hline
Real                     & 0.506          & 0.800         & 2.986                   & -           & -          & -                        & -                      & -                          \\ \hline
MotionGPT~\cite{jiang2023motiongpt}                & \textbf{0.534}          & 0.803         & \underline{2.978}                   & 42.61       & 6.04       & 34.47                    & \underline{7.92}                   & 31.57                      \\
TM2T~\cite{guo2022tm2t}                     & \underline{0.525}          & \textbf{0.814}         & 2.995                   & \textbf{61.76}       & \textbf{21.98}      & \textbf{47.40}                    & \textbf{71.12}                  & \underline{37.27}                      \\ \hline
Ours                     & 0.520          & \underline{0.806}         & \textbf{2.958}                   & \underline{52.84}       & \underline{9.27}       & \underline{40.11}                    & 6.22                   & \textbf{40.90}                      \\ \hline \hline
\end{tabular}
}}
\vspace{-3mm}
\caption{\small Comparison of motion captioning on HumanML3D dataset.}
\label{table:motion-to-text}
\vspace{-6mm}
\end{table*}

\noindent\textbf{Text-to-Music.}
In Table~\ref{table:text-to-music}, we compare our UniMuMo with Riffusion~\cite{Forsgren_Martiros_2022}, Mubert~\cite{mubert2022}, Mousai~\cite{schneider2023mo}, MusicLM~\cite{Agostinelli2023MusicLMGM}, MusicGen~\cite{copet2023simple} and AudioLDM 2~\cite{liu2023audioldm2}. We evaluate the performance on MusicCaps, with results of SOTAs directly sourced from their respective papers. We employ three metrics: Frechet Audio Distance (FAD$_{\text{VGG}}$)~\cite{kilgour2018fr}, Kullback-Leibler Divergence (KL)~\cite{kreuk2022audiogen} and CLAP similarity (CLAP)~\cite{wu2023large, huang2023make}. The first two metrics measure the audio quality, while the last one measures the correspondence between generated audio and text descriptions. 
Note that the audio quality of our model does not match with SOTA models. We argue that this might be due to the poor audio quality of our training data. Following MusicGen, we also use vocal-free training data. To achieve this, we use Demucs~\cite{defossez2021hybrid,rouard2022hybrid} to remove the vocal part of the music in Music4All dataset. Nonetheless, we observe that many of the processed audio are of bad quality. This is testified by the experiment of fine-tuning MusicGen on our dataset for the same number of epochs while keeping all other settings the same (\eg, sequence length, batch size). As shown in Table~\ref{table:text-to-music}, the audio quality of the tuned model also degrades. We also tried training the model on the original dataset with vocals, resulting in improved quantitative scores. However, the generated music is not perceptually good, often filled with weird and meaningless vocals.  This phenomenon, where training on music with vocals yields better quantitative scores, is also reported in MusicGen.

\noindent\textbf{Dance-to-Music.}
In Table~\ref{table:motion-to-music}, we compare UniMuMo with  Dance2Music~\cite{aggarwal2021dance2music}, Foley Music~\cite{gan2020foley}, CMT~\cite{di2021video}, D2M-GAN~\cite{zhu2022quantized} and CDCD~\cite{zhu2022discrete} on dance-conditioned music generation.
For evaluation, we adopt Beats Coverage and Beats Hit~\cite{zhu2022quantized}, both of which measure the alignment of generated music with motion. 

\noindent\textbf{Music/Text-to-Dance.}
In Table~\ref{table:motion_generation}, we compare UniMuMo's dance-generation capabilities  with Bailando~\cite{siyao2022bailando}, FACT~\cite{li2021ai} and EDGE~\cite{tseng2023edge} on AIST++ dataset. 
We evaluate UniMuMo on both music-conditioned and text-conditioned dance generation tasks. Although there is currently no established benchmark for the text-to-dance task, we can also apply the same evaluation metrics to measure and compare the quality of generated dance. For evaluation metrics, we adopt kinetic distribution spread (Dist$_{\text{k}}$) and geometric distribution spread (Dist$_{\text{g}}$) to measure the diversity. Additionally, we employ the beat alignment score to measure the alignment between conditioning audio and generated dance. Following EDGE, we evaluate the motion sequences on 5-second clip. For text-to-dance, we directly evaluate the dance that is jointly generated with music, conditioned on both music and motion captions, and we calculate the beat alignment score between  the generated dance and music.
The quantitative scores show that UniMuMo achieves competitive results on music-conditioned dance generation, even though it hasn't been fine-tuned on AIST++ music. For text-conditioned generation, it achieves inferior dance quality since there is no ground truth music for reference, but also gains a higher beat alignment score due to the joint generation.

\noindent\textbf{Music-to-Text.}
In Table~\ref{table:music-to-text}, we compare UniMuMo against SOTA music captioning models including LTU~\cite{gong2023listen}, LP-MusicCaps~\cite{doh2023lp} and MU-LLaMA~\cite{liu2023music}. The evaluation is conducted on the MusicQA dataset released by~\cite{liu2023music}, which is a music-related question-answering dataset. We take the answers to the question "Describe the audio" together with the corresponding music as evaluation data, totaling 552 music-caption pairs.
Following MU-LLaMa, the metrics we use includes Bleu, Meteor, Rouge$_\text{L}$ and BertScore, which are all common evaluation metrics in natural language processing. 

\noindent\textbf{Motion-to-Text.}
In Table~\ref{table:motion-to-text}, we compare UniMuMo with TM2T~\cite{guo2022tm2t} and MotionGPT~\cite{jiang2023motiongpt} for motion captioning using the HumanML3D test set. 
Following MotionGPT, we adopt the motion-retrieval precision (R-Precision) to measure the accuracy of motion-text matching using top-1 and top-3 retrieval accuracy, multi-modal distance (MM Dist) to measure the distance between motion and text, and other popular natural language processing metrics, including Blue, Rouge, Cider and BertScore, to assess the linguistic quality. Since we source only 50\% of our training motion data from HumanML3D, and the motion is augmented to align with music beats, UniMuMo still lags behind the best SOTA in certain metrics for HumanML3D motion captioning task.

Based on the quantitative results presented above, UniMuMo achieves competitive performance compared to the SOTA benchmarks across various single-modal generation tasks. Specifically, in the motion-to-music, music-to-motion, music captioning and motion captioning tasks, UniMuMo generally ranks second among the SOTAs. However, in the text-to-music task, UniMuMo's performance is not as competitive, which we argue may be attributed to the limitations in our training data.

\subsection{Ablation Studies}

\begin{table}[htbp]
\small
\centering
\setlength{\tabcolsep}{3pt}
\resizebox{\linewidth}{!}{
\begin{tabular}{l|ccc|ccc}
\hline \hline
           & FAD$_{\text{VGG}}\downarrow$  & KL$\downarrow$   & CLAP$\uparrow$ & Dist$_{\text{k}}\rightarrow$ & Dist$_{\text{g}}\rightarrow$ & Beat Align.$\uparrow$ \\ \hline
Full       & 5.93 & 1.99 & 0.27 & 10.54   & 8.15    & 0.24        \\ \hline
Ablation 1 & 6.75 & 2.13 & 0.26 & 6.03    & 8.28    & 0.22        \\
Ablation 2 & 6.79 & 2.06 & 0.26 & 6.73    & 7.35    & 0.23        \\
Ablation 3 & 6.29 & 2.11 & 0.24 & 6.03    & 8.28    & 0.22        \\
Ablation 4 & 6.41 & 2.06 & 0.27 & 9.22    & 6.58    & 0.22        \\
Ablation 5 & 7.10 & 2.22 & 0.23 & 9.61    & 9.01    & 0.23           \\ \hline \hline
\end{tabular}
}
\vspace{-3mm}
\caption{\small Comparisons of our full model with different ablation studies on MusicCaps for music generation and our Music4All for dance generation. 
Ablation 1-2 show the results of using an independent motion VQVAE for encoding motion sequences. Ablation 3 shows the results of model without the key structures of separate embedder and MoE. Ablation 4 shows the results of using a mixture of training tasks during training. Ablation 5 shows the result of training our model from scratch. }
\label{table:ablation}
\end{table}

\vspace{-4mm}
In the ablation study, we first evaluate the effectiveness of the proposed joint codebook encoding by training our model with motion encoded by an independent motion VQ-VAE (ablation 1 and 2). We then assess the impact of the additional separate embedder and FFN introduced to MusicGen by training the model without them (ablation 3). We also investigate training the model with multiple tasks, rather than the single proposed music-motion joint generation (ablation 4). Finally, we evaluate the effectiveness of using a pre-trained model for initialization by training the model from scratch (ablation 5). All ablations are compared with our benchmarks on MusicCaps for music generation and Music4All for motion generation, as shown in Table~\ref{table:ablation}. More details and analysis are in Appendix~\ref{appendix:ablation_analysis}.

%% file: 10_conclusion.tex
\section{Conclusion}
\label{sec:conclusion}
\vspace{-1mm}
In this paper, we introduce UniMuMo, the first unified framework for arbitrary generation across music, motion, and text. To address the limitations of paired multimodal data, we expand existing datasets with rhythm-based music-motion alignment and text augmentation, thus creating a comprehensive new dataset. To build a unified model, we propose novel architectural designs, including a music-motion joint tokenizer for bridging modality gaps and a music-motion parallel generation scheme for synchronized music and motion generation. Extensive experiments show that UniMuMo achieves competitive performance in all unidirectional generative tasks. We believe our framework will not only open up new avenues for multimodal generation but also inspire future advancements in this rapidly evolving field.

%% file: 12_appendix.tex
\noindent \textbf{\large Appendix}

\section{Visual Beats Detection Details}
\label{appendix:visual beat detections}
As discussed in the main paper, there are three steps to obtain visual beats from 3D motion sequence $M \in \mathbb{R}^{T_m \times J \times 3}$ where $T_m$ is the number of frames, $J$ is the number of joints, and the third dimension represents $x,y,z$ coordinates. Inspired by the idea used to find visual impacts in raw videos~\cite{davis2018visual}, we adapt this approach to find local saliency in motion sequences. The main idea is to use the sudden visible deceleration of the motion sequence as the basis of the heuristic to find ``visual beats''. The three steps have very similar physical meanings to traditional music beat detection techniques~\cite{bock2013maximum, ellis2007beat}.

In the first step, we compute the motion directogram $M_d$, a 2D matrix that factors motion into different motion angles, as described in the main paper. The motion directogram is similar to the audio spectrogram, which could offer spectral flux to measure the change in amplitude of different frequencies over time. Therefore, in the second step, we compute the pre-direction deceleration of $M_d$ as an analogue for spectral flux to obtain the motion flux $M_f$. In audio, the onset envelope could be inferred from the spectral flux. In motion, we obtain the visual impact envelope, called kinematic offsets $M_k$, by averaging each frame of $M_f$ and filtering the top $1\%$ peaks. With the onset envelope, usually, an onset detection followed by dynamic programming-based beat tracking algorithms~\cite{bock2013maximum, ellis2007beat} are used to find the most likely periodic music beats. In motion sequence, we also use dynamic programming to compute the visual beats by designing an objective function that selects strong visual changes from kinematic offsets and encourages equal-spacing beats. We optimize the objective function:
\begin{align*}
V(\mathbf{m}) = \sum_{j=1}^{n} u(m_j) + \alpha \sum_{j=1}^{n-1} V_T(m_j, m_{j+1})\\
 V_T(m_j, m_{j+1}) = \frac{T[\text{bin}(m_{j+1} - m_{j})]}{T_{max}} - 1.0
\end{align*}
where $u$ is the kinematic offset value of the candidate beat to encourage strong visual impacts, $\alpha$ is the weight to balance the two terms, $T$ is the autocorrelation mean over the local time window, and $\{m_j\}^n_{j=1} \in \mathbf{m}$ is a subset of candidate beats. The $V_T(m_j, m_{j+1})$ regularizes the estimated tempos within a local window to encourage equal-spacing beats. We measure the deviation by computing the time-dependent autocorrelation function $T$ on kinematic offsets.

\section{Music Motion Alignment Details}
\label{appendix:music motion alignment}
We apply dynamic time warping (DTW) to compute the optimal alignment between music beats $B_m$ and visual beats $B_v$. The optimal alignment minimizes the sum of distances between aligned elements, even though the lengths of $B_m$ and $B_v$ may differ. The local distance between elements of $B_m$ and $B_v$ is computed by Euclidean distance. Regarding the transitions allowed while searching for the minimum-distance path, we use the Rabiner-Juang step pattern~\cite{rabiner1999fundamentals}. We use python-dtw package~\cite{giorgino2009computing} to find the alignment. Finally, we warp motion sequences according to the warping curve.

\section{Text Description Construction Details}
\label{appendix:text constructions}

\begin{table*}[htbp]
\centering
\small
\resizebox{\textwidth}{!}{
\begin{tabular}{l|l}
\hline \hline
Description Type   & Examples                                                                                                                                                                                                                                                                                                                                                                                                                                                                                                                                                         \\ \hline \hline
ChatGPT Generated  & \begin{tabular}[c]{@{}l@{}}Blending elements of underground hip-hop, ekip, and rap, this music exudes a \\ gentle yet quick energy that immerses the listener in its unique category.\\ This high-energy song combines the raw power of indie rock with the smooth and expressive \\ elements of jazz, resulting in a unique and captivating musical experience.\\ This high-tempo track seamlessly combines German pop and alternative styles, boasting a \\ spirited and lively atmosphere that will keep listeners engaged from start to finish.\end{tabular} \\ \hline 
MU-LLaMa Generated & \begin{tabular}[c]{@{}l@{}}The music is a blend of neofolk, martial industrial, and dark ambient.\\ The audio is a progressive rock/metal song with a fast tempo, steady drumming, and a bass guitar rhythm.\\ The music is a bossa nova/jazz/MPB/soul fusion with a touch of Brazilian rhythms.\end{tabular}                                                                                                                                                                                                                                                    \\ \hline 
Motion Desctiption & \begin{tabular}[c]{@{}l@{}}The style of the dance is ballet jazz.\\ This is a break style dance.\\ The genre of the dance is LA style hip-hop.\end{tabular}                                                                                                                                                                                                                                                                                                                                                                                                      \\ \hline \hline
\end{tabular}}
\vspace{1mm}
\caption{Examples of three kinds of synthesized text descriptions.}
\label{table:description examples}
\end{table*}

\subsection{Music Caption Generation with Music Language Model}
We employed MU-LLaMa~\cite{liu2023music}, a specialized large language model for music-related Q\&A tasks, to generate music descriptions using genre metadata from the Music4All dataset. Our experiments show that MU-LLaMa  generally understands music effectively, accurately assesses tags, and integrates them into cohesive descriptions. However, we also notice that the generated descriptions often lack diversity both in sentence structure and content. Repeated results are also observed.

\subsection{Music Caption Generation with ChatGPT and Template Filling}
We adopt energy, tempo, genres, and tags from the metadata of Music4All~\cite{santana2020music4all}. Energy, indicating musical intensity and activity, ranges from 0.0 to 1.0. Tempo, measured in beats per minute (BPM), reflects the music's speed. The dataset categorizes music into 853 unique genres and includes 19,541 tags.

First, we convert energy and tempo into descriptive tags with the criteria shown in Table~\ref{table:tempo_tag} and~\ref{table:energe_tag}. Adverbs and adjectives are randomly selected and paired, and the thresholds are manually determined by listening to samples and comparing values. Then, we construct phrases from tags using random templates. We construct a tempo phrase from tempo description, an energy phrase from energy description and a tag phrase from genres and tags. The choices of templates are shown in Table~\ref{table:phrase construction}. Next, we randomly shuffle, dropout and concatenate the phrases to construct raw music text descriptions. Finally, we refine the descriptions with ChatGPT~\cite{brown2020language}. We set the ChatGPT content as ``You are an expert in music, skilled in writing music comments and descriptions.'' and use the prompt ``Here are \{n\} music descriptions, please polish them separately into fluent and meaningful sentences with details. Please return the polished results in the format of ``1: content... 2: content... ..."'' to polish n descriptions on each request.

This approach ensures that our synthesized descriptions are more natural than direct tag concatenation and more diverse than using full descriptive templates. Moreover, it has more control on the synthesized results than directly asking ChatGPT to write descriptions from tags. However, since the generated captions source the musical information only from genre, tag, intensity, and tempo, they often cannot provide specific details such as the instrument composition or the emotional tone. Additionally, the descriptive tags of energy and tempo can sometimes be imprecise due to the inaccuracies of metadata. Therefore, during training we leverage a mixture of the above two methods for music captioning to strike a balance between musical accuracy and diversity.

\begin{table}[]
    \small
  \centering 
  \resizebox{\linewidth}{!}{
  \scalebox{0.8}{
  \begin{tabular}{l|l|l}
    \hline \hline
    Range                                & Adverb                                                             & Adjective                                                                                               \\ \hline \hline
    tempo $<$ 60                   & \begin{tabular}[c]{@{}l@{}}extremely, \\ very\end{tabular}         & \begin{tabular}[c]{@{}l@{}}slow, languid,\\ lethargic, relaxed,\\ leisure, chilled\end{tabular}         \\ \hline
    60 $\leq$  tempo $<$ 75   &                                                                    & \begin{tabular}[c]{@{}l@{}}slow, languid, \\ lethargic, relaxed, \\ leisure, chilled\end{tabular}       \\ \hline
    75 $\leq$  tempo $<$ 110  &                                                                    & \begin{tabular}[c]{@{}l@{}}moderate, easy-going, \\ laid-back,medium, \\ balanced, neutral\end{tabular} \\ \hline
    110 $\leq$  tempo $<$ 150 &                                                                    & \begin{tabular}[c]{@{}l@{}}fast, upbeat, high, \\ brisk, quick, rapid, \\ swift\end{tabular}            \\ \hline
    tempo \textgreater{}= 150            & \begin{tabular}[c]{@{}l@{}}extremely, \\ very, highly\end{tabular} & \begin{tabular}[c]{@{}l@{}}fast, upbeat, high, \\ brisk, quick, rapid, \\ swift\end{tabular}            \\ \hline \hline
    \end{tabular}
  }
  }
  \vspace{-3mm}
  \caption{Choices of descriptive tags for tempo. Adverbs and adjectives are randomly chosen and paired.} 
  \label{table:tempo_tag}
\end{table}

\begin{table}[]
    \small
  \centering 
  \resizebox{\linewidth}{!}{
  \scalebox{0.8}{
  \begin{tabular}{l|l|l}
    \hline \hline
    Range                                 & Adverb                                                             & Adjective                                                                                                         \\ \hline \hline
    energy $<$ 0.1                  & \begin{tabular}[c]{@{}l@{}}extremely, \\ very\end{tabular}         & \begin{tabular}[c]{@{}l@{}}soft, calm, peaceful, \\ serene, gentle, light, \\ tranquil, mild, mellow\end{tabular} \\ \hline
    0.1 $\leq$  energy $<$ 0.4 &                                                                    & \begin{tabular}[c]{@{}l@{}}soft, calm, peaceful,\\ serene, gentle, light,\\ tranquil, mild, mellow\end{tabular}   \\ \hline
    0.4 $\leq$  energy $<$ 0.7 &                                                                    & \begin{tabular}[c]{@{}l@{}}moderate, comfortable,\\ balanced, relaxing\end{tabular}                               \\ \hline
    0.7 $\leq$  energy $<$ 0.9 &                                                                    & \begin{tabular}[c]{@{}l@{}}intense, powerful, strong,\\ vigorous, fierce,\\ potent, energetic\end{tabular}        \\ \hline
    energy \textgreater 0.9               & \begin{tabular}[c]{@{}l@{}}extremely, \\ very, highly\end{tabular} & \begin{tabular}[c]{@{}l@{}}intense, powerful, strong,\\ vigorous, fierce,\\ potent, energetic\end{tabular}        \\ \hline \hline
    \end{tabular}
  }}
  \vspace{-3mm}
  \caption{Choices of descriptive tags for energy.} 
  \label{table:energe_tag}
\end{table}

\begin{table}[h]
\centering
\small
\resizebox{0.38\textwidth}{!}{
\begin{tabular}{l|l}
\hline \hline
Phrase Type   & Phrase Choices                                                                                                                                                                                                                                 \\ \hline \hline
Tempo Phrase  & \begin{tabular}[c]{@{}l@{}}with a $< >$ tempo,\\ whose speed is $< >$,\\ a $< >$ music,\\ set in a $< >$ pace\end{tabular}                                               \\ \hline
Energy Phrase & \begin{tabular}[c]{@{}l@{}}which is $< >$,\\ with $< >$ intensity,\\ a $< >$ music,\\ whose energy is $< >$\end{tabular}                                               \\ \hline
Tag Phrase    & \begin{tabular}[c]{@{}l@{}}This is atrack which is $< >$,\\ This song has the style of $< >$,\\ The music is $< >$,\\ The genre of the music is $< >$\end{tabular} \\ \hline \hline
\end{tabular}}
\caption{Choices of phrase template for tempo, energy and tags. Inside $< >$, we fill in tempo tag, energy tag or a list of genres and tags.}
\label{table:phrase construction}
\end{table}

\subsection{Motion Caption Generation}
The only metadata that is available for AIST++~\cite{li2021learn} and DancedDB~\cite{AMASS:ICCV:2019} is a genre tag.  Therefore, we directly construct the motion descriptions using the following templates: ``The genre of the dance is $< >$", ``The style of the dance is $< >$.", ``The is a $< >$ style dance.", where $< >$ is filled with the genre tag.

Finally, as mentioned in the paper, we apply text conditioning separately to music and motion, adding classifier-free guidance dropout independently. Examples of the several types of descriptions mentioned above are shown in Table~\ref{table:description examples}.

\section{Illustration of Music Motion Parallel Generation}
\label{appendix:parallel generation}

During the training of UniMuMo, we introduce a music motion parallel generation scheme, which not only allows for the joint generation of music and motion, but also enables zero-shot music-to-motion and motion-to-music generation. In this section, we provide illustrations to further explain  the training and inference processes. 

\begin{figure}[htbp]
    \centering
    \begin{subfigure}[b]{0.45\textwidth}
        \includegraphics[width=\textwidth]{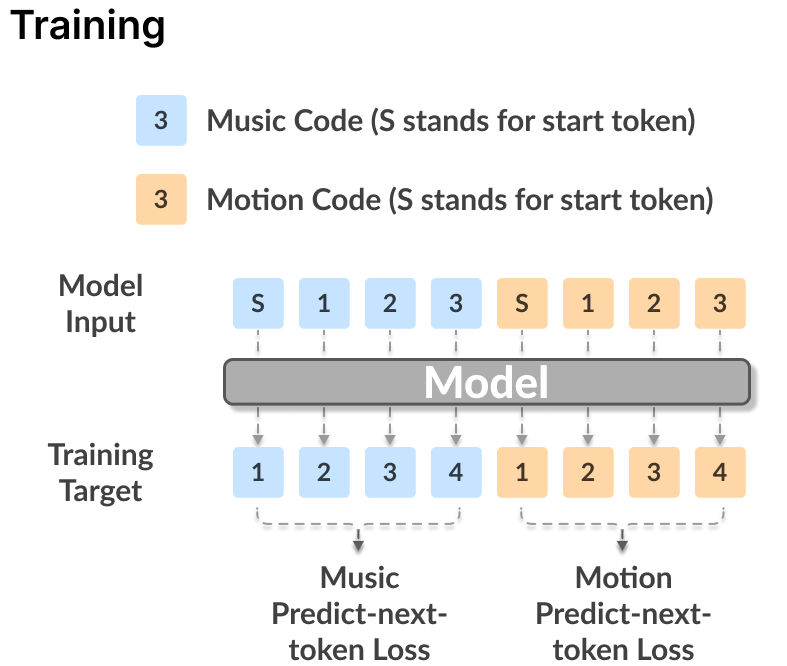}
        \caption{An illustration of one forward pass during UniMuMo's training. Music tokens (including the music start token) are denoted in blue blocks and motion tokens (including the motion start token) are denoted in orange blocks. The numbers in the block denote the timestep of each token.}
        \label{fig:parallel_training}
    \end{subfigure}
    \hfill 
    \begin{subfigure}[b]{0.45\textwidth}
        \centering
        \includegraphics[width=0.7\textwidth]{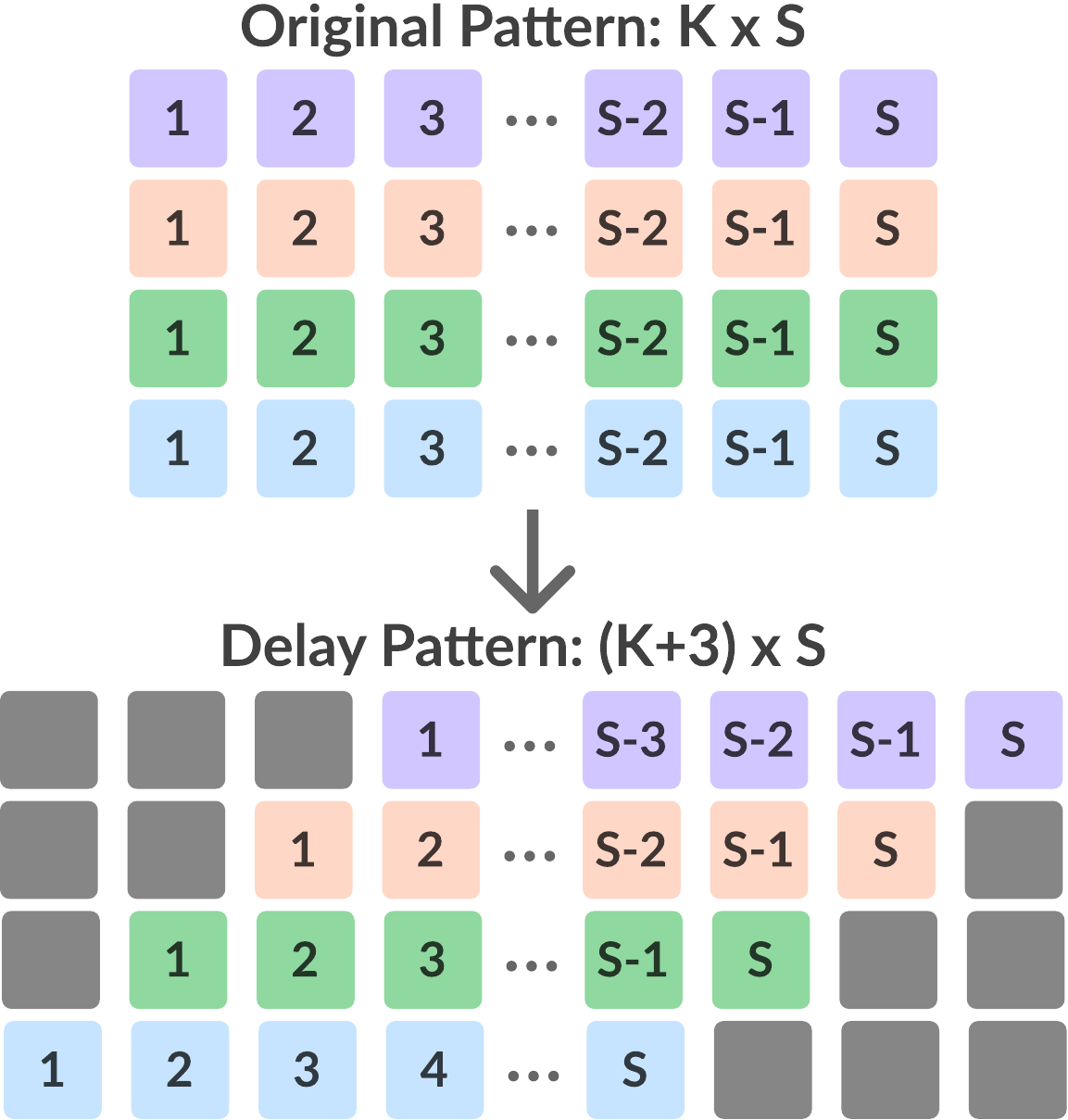}
        \caption{An illustration of the delay pattern in MusicGen. Each color represents a different layer of the residual codebook, and the numbers on the blocks indicate the timestep. After applying the delay pattern, the tokens denoted in grey are padded with a special empty token.}
        \label{figs:delay_pattern}
    \end{subfigure}
    \caption{Illustrations on the technical details in our training process.}
\end{figure}

During training, as illustrated in Figure~\ref{fig:parallel_training}, music and motion are trained on language modeling task separately. In each forward pass, the system is trained to predict the next music token for music and the next motion token for motion. The predict-next-token losses are calculated separately for each modality and then summed up with different weights. The customized cross-modal causal-attention mask, each quarter of which is a lower triangular matrix, ensures that each modality can causally attend to itself while also allowing both modalities can causally attend to each other.

\begin{figure}[htbp]
    \centering
    \begin{subfigure}[b]{0.45\textwidth}
        \includegraphics[width=0.8\textwidth]{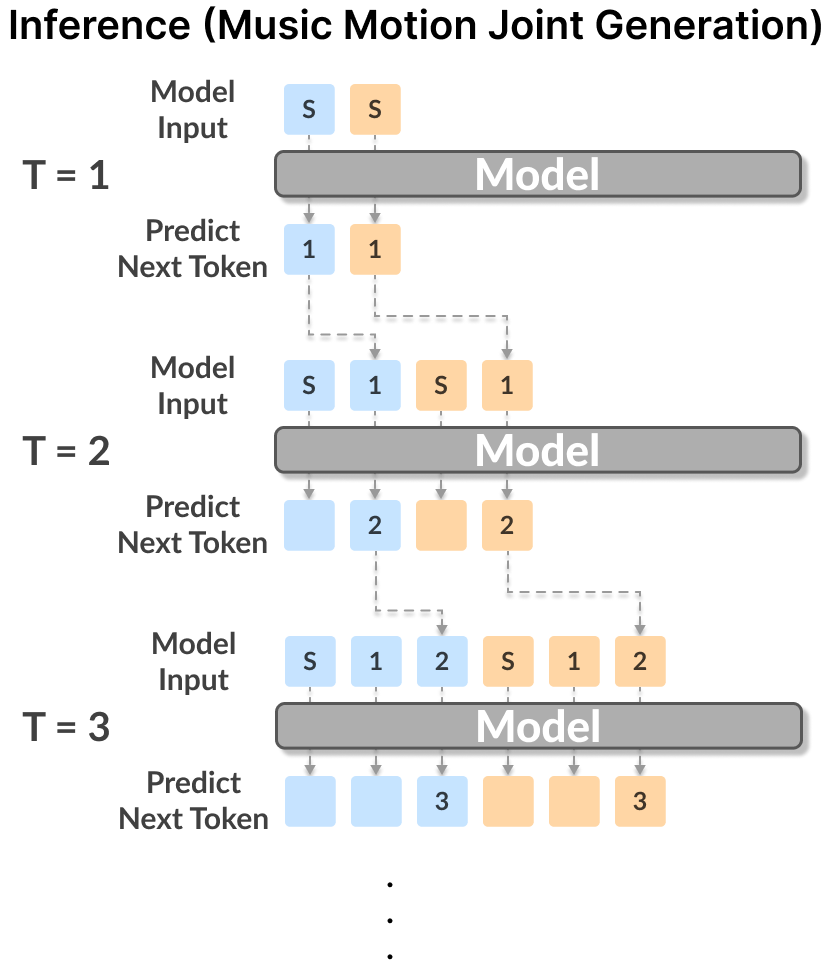}
        \caption{An illustration of how UniMuMo conduct music motion parallel generation. In each timestep T, one forward pass is performed.}
        \label{fig:parallel_infer}
    \end{subfigure}
    \hfill 
    \begin{subfigure}[b]{0.45\textwidth}
        \includegraphics[width=\textwidth]{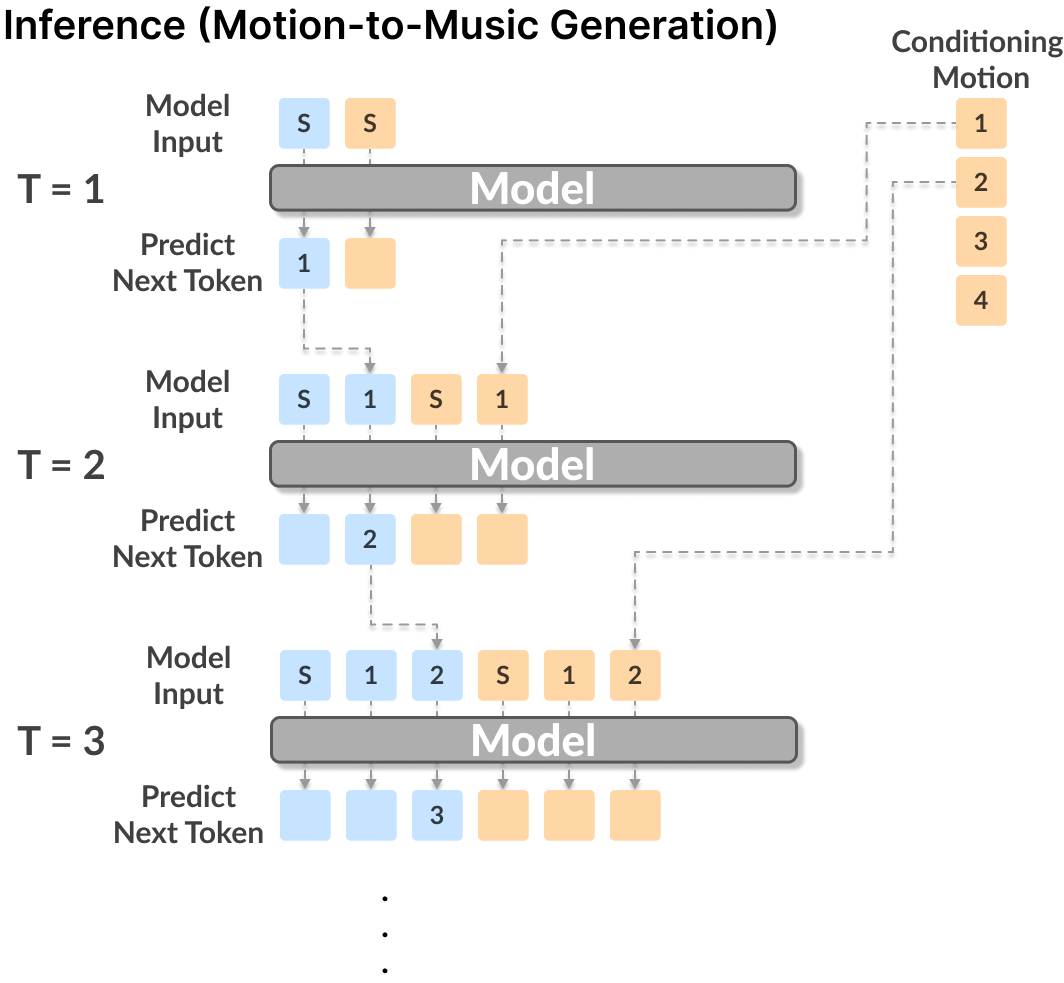}
        \caption{An illustration of UniMuMo's motion-to-music generation. The music-to-motion generation is similar to this process.}
        \label{fig:parallel_infer_single_modal}
    \end{subfigure}
    \caption{Illustrations on the technical details in the inference process.}
\end{figure}

Note that in our implementation, music and motion are encoded using a four-layer residual codebook, resulting four tokens at each timestep. As mentioned in the paper, we follow the approach of MusicGen~\cite{copet2023simple}, adopting a delay pattern to transform the input tokens of shape $K \times S$, where $K$ is the number of codebooks, chosen to be 4 in our implementation, and $S$ is the length of the sequence of tokens. Figure~\ref{figs:delay_pattern} illustrates this pattern, with more details available in their original paper. During implementation, we first transform the music tokens and motion tokens separately according to this pattern, and then concatenate them to form the model input. 

During inference, as shown in Figure~\ref{fig:parallel_infer} and~\ref{fig:parallel_infer_single_modal}, one music token and one motion token are sampled on each forward pass. For joint music-motion generation, the sampled next music token and next motion token are placed in the corresponding positions of the input for the next timestep. When generating music conditioned on motion, the conditioning motion token at the same timestep is used instead of the predicted next motion token.

\section{User Study on Music-Motion Alignment}
\label{appendix:user_study}
\begin{figure}
    \centering
    \includegraphics[width=\linewidth]{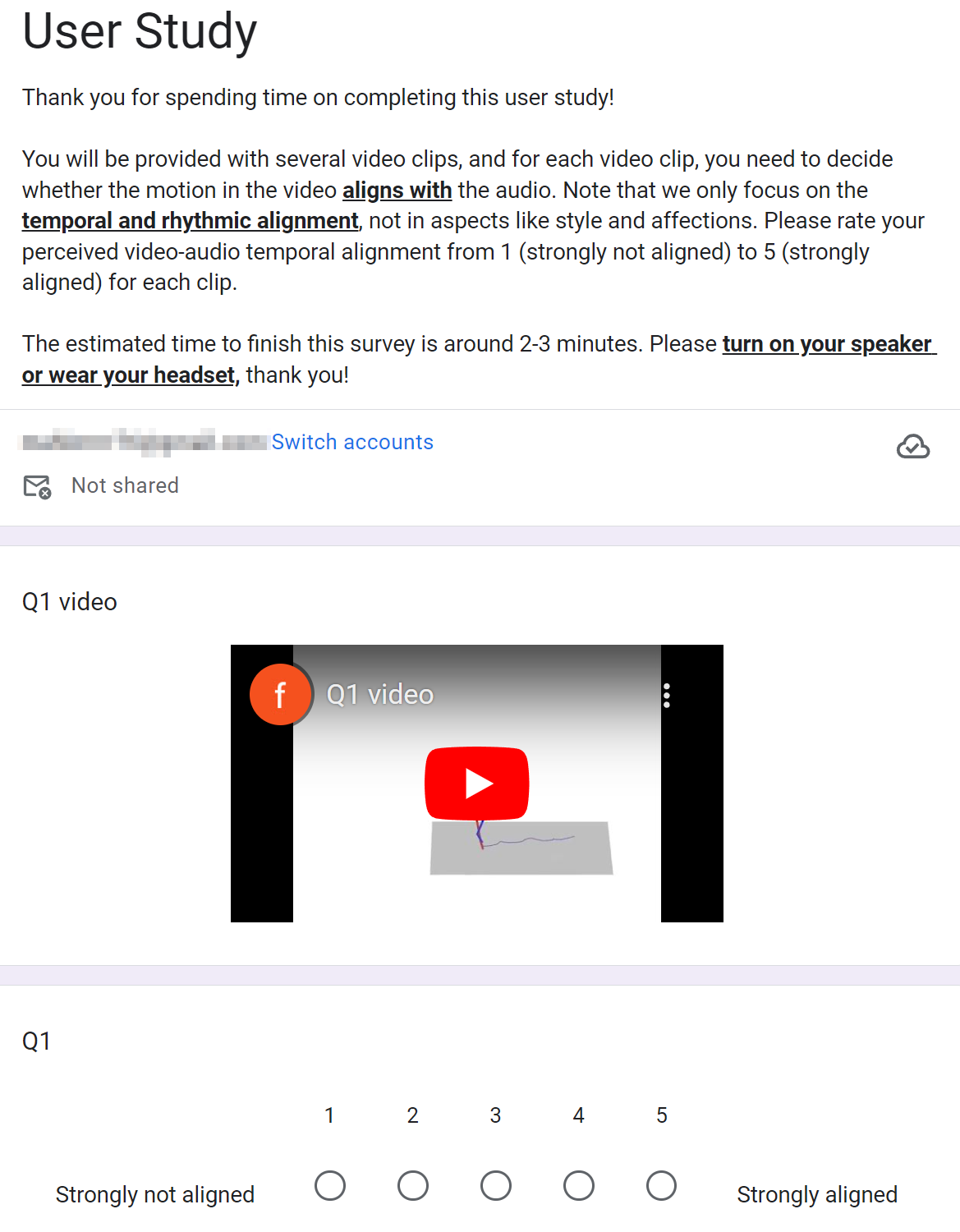}
    \caption{A screen shot of the user study form for evaluating our music-motion alignment algorithm.}
    \label{fig:user-study}
\end{figure}

Since no ground truth for measuring the accuracy of the music-motion alignment, we further conduct a user study, where users are required to rate the audio-visual alignment from 1 (not aligned) to 5 (aligned) based on their perception. According to the responses of 8 users, each presented with 20 videos (half aligned, half randomly paired), the average score is 3.95 for aligned results and 3.26 for random pairs, demonstrating the algorithm's effectiveness. A screenshot of the survey form is in Figure~\ref{fig:user-study}.

\section{Implementation Details}
\label{appendix:evaluation details}

\subsection{Training}
\noindent\textbf{Datasets.}
For the music training dataset, we mainly use Music4All~\cite{santana2020music4all}, which consists of 109K 30-second soundtracks with corresponding metadata, such as genre, energy, and tempo. 
For evaluation and comparison with prior work on music generation, we use the MusicCaps benchmark~\cite{Agostinelli2023MusicLMGM}, which is composed of 5.5K ten-second samples with expert-prepared text descriptions. For evaluation on music captioning, we leverage the MusicQA Dataset~\cite{liu2023music}, which contains 560 music tracks with textual descriptions. 

For motion dataset, we use a mixture of 3D dance dataset AIST++~\cite{li2021learn}, DancedDB~\cite{AMASS:ICCV:2019} and 3D motion dataset HumanML3D~\cite{Guo_2022_CVPR}. AIST++ contains 311 minutes of dance across 30 subjects and 10 genres, while DanceDB contains 203.38 minutes of dance across 20 subjects. Text descriptions for both datasets are synthesized by filling templates with metadata. HumanML3D, a much larger dataset of general motion, contains 14616 motions with 44970 text descriptions, in total 28.59 hours long. We upsample its motion data from 20 Hz to 60 Hz to align with the frame rate of the AIST++ dataset.

\noindent\textbf{Data Preparations.}
We remove the vocal part of all the training audios using Demucs~\cite{defossez2021hybrid,rouard2022hybrid} and keep only the instrumental part.
We randomly pair each soundtrack with 5 motion sequences, with approximately 50\% drawn from 3D dance datasets AIST++ and DanceDB, and the remainder from HumanML3D. The pairing and tokenization of music and motion are done in advance to save training time. 

\noindent\textbf{Stage 1: Music and Motion Tokenization Model.}
We adopt the default Encodec from the MusicGen model, which compresses 32K Hz waveform into 50 Hz. The RVQ has 4 codebooks, each with 2048 entries of dimension 128. The motion encoder encodes 60 Hz, 263-dimensional motion features into 50 Hz, 128-dimensional features suitable for quantization by the RVQ.
The motion encoder-decoder, together with the frozen RVQ, is trained on 2-second motion data, with a batch size of 336 and a learning rate of 2e-4. 

\noindent\textbf{Stage 2\: Music-Motion Decoder Model.}
In stage 2, the model we fine-tune is MusicGen-small, a 300M transformer decoder model together with a 120M T5 text encoder. We train the model on 10-second aligned music-motion pairs with a batch size 144. We train the model for 15K steps with AdamW optimizer~\cite{loshchilov2017decoupled}, $\beta_1 = 0.9$, $\beta_2 = 0.95$, and a learning rate of 5e-5. The training takes around 6 hours on 48 Tesla V-100 32G GPUs. 

\noindent\textbf{Stage 3: Music-Motion Captioner Model.}
In stage 3, we choose the T5-base model as the text decoder. We train the model on 10-second unpaired music and motion sequences. Motion sequences shorter than this duration are zero-padded to 10 seconds. The other settings remain the same as in stage 2.

\subsection{Evaluation}
\noindent \textbf{Text-to-Music.}
The evaluation scripts for FAD and KL are directly from the evaluation repository of AudioLDM\footnote{{\text{https://github.com/haoheliu/audioldm\_eval}}}~\cite{liu2023audioldm}. We use the default pre-trained model in the official webpage of CLAP\footnote{{https://github.com/LAION-AI/CLAP}} to calculate the CLAP score.  All the scores for SOTAs are directly borrowed from the corresponding papers, except for MusicGen. As explained on their Hugging Face pages, their publicly released models are trained on another set of vocal-free music, resulting in slightly lower quantitative scores. Since we directly fine-tune their released model, we report the scores as presented on the Hugging Face page instead.

\noindent \textbf{Motion-to-Music.}
The evaluation scripts for Beats Coverage and Beats Hit are directly sourced from D2M-GAN~\footnote{{https://github.com/L-YeZhu/D2M-GAN}}~\cite{zhu2022quantized}. The test music-motion pairs are 2 seconds long, and the segmentation is also from D2M-GAN~\cite{zhu2022quantized}. All  scores for other models are directly borrowed from CDCD~\cite{zhu2022discrete}.

\noindent \textbf{Music/Text-to-Dance.}
We directly use the script in Bailando~\footnote{{https://github.com/lisiyao21/Bailando}}~\cite{siyao2022bailando} for evaluating feature distributions (Dist$_{\text{k}}$ and Dist$_{\text{g}}$). For the beat alignment score, we modify the script from~\cite{siyao2022bailando} to use the Librosa~\cite{mcfee2015librosa} API for music beat detection. All scores for other models are sourced from~\cite{tseng2023edge}. Regarding the choice of evaluation metrics, we did not include the Physical Foot Contact score (PFC) proposed in~\cite{tseng2023edge}, as their provided script does not yield correct scores on our data, even when evaluating the ground truth. The discrepancy may be due to the slight differences in motion representation (\eg, fps, number of joint). 
According to~\cite{tseng2023edge}, the Frechet Inception Distances on kinetic features and geometric features are also not employed, as they are not considered reliable for measuring dance quality. For testing data, we use the original test set split from the AIST++ dataset, and slice the dance-music pairs into 5-second segments using the script from~\cite{tseng2023edge}.

\noindent\textbf{Music-to-Text.}
We use the evaluation script from MU-LLaMa~\footnote{{https://github.com/shansongliu/MU-LLaMA}}~\cite{liu2023music} for the testing metrics. The testing dataset, MusicQA, is also publicly available in~\cite{liu2023music}. As we only take the subset of the dataset related to music captioning as our test set, we re-evaluate all previous SOTAs and report their results.

\noindent\textbf{Motion-to-Text.}
We adopt the evaluation script from the open-sourced code of MotionGPT~\footnote{{https://github.com/OpenMotionLab/MotionGPT}}~\cite{jiang2023motiongpt}. Since they announced that they used a different package for NLP-related metrics calculation due to package conflict, we re-evaluated their model and TM2T~\cite{guo2022tm2t} on our HumanML3D test set. This accounts for the differences between our reported results and theirs.

\section{More Analysis on Ablation Study}
\label{appendix:ablation_analysis}

\noindent \textbf{Ablation 1-2: Effect of joint codebook encoding.} 
In this analysis, we examine the effectiveness of using a shared codebook that maps motion into the same feature space as music. Specifically, we train the model on motion codes extracted by an independent motion VQ-VAE. In ablation 1, we don't initialize the motion embedder with corresponding pre-trained weights in MusicGen, while in ablation 2, we perform such initialization, which is also the practice in our full model. The results of both ablation studies yield inferior scores, demonstrate the effectiveness of our method. Importantly, ablation 2 proves that it is not merely the initialization of the motion embedders that facilitates training, but it is only when the initialized motion embedders are input with meaningful motion tokens that training is facilitated.

\noindent \textbf{Ablation 3: Effect of additional architectures.}
We also examine the effectiveness of the additional components we introduce to the pre-trained MusicGen, specifically the separate motion embedder and Mixture of Expert (MoE) structure. 
In ablation 3, we train the model without MoE and the separate motion embedder, allowing music and motion to share most of the MusicGen model, except the linear classifier at the end. Experimentally, this results in degraded performance, especially in motion generation. 

\noindent \textbf{Ablation 4: An alternative multi-task training scheme.}
In stage two, we employ a single training task of music-motion joint generation. However, within this parallel transformer structure, it is also intuitive to adopt a mixture of three training tasks: music-motion generation, music-to-motion and motion-to-music. We explore this idea in ablation 4, where we randomly select among the three tasks during training and apply the corresponding cross-modal self-attention masks (\eg, in music-to-motion, we allow music to only causally attend to itself, and motion to causally attend to itself and fully attend to music). However, the results are not satisfactory. We hypothesize that the reasons could be 1) a mixture training tasks might result in gradient conflicts; 2) the difficulty level of the three tasks varies, so simply randomly selecting the task might not be sufficient; and 3) the rhythms of music and dance have regular patters, so they might not necessarily need to be fully attended to when performing cross attention. 

\noindent\textbf{Ablation 5: Effect of using pre-trained model.}
In ablation 5, we train the same model from scratch, keeping other settings unchanged. The best results shown in the table are achieved after around 30K iterations of training, compared with only 15K iterations of training if the pre-trained model is loaded.